\documentstyle[aps,epsfig]{revtex}

\def \be {\begin{equation}}
\def \ee {\end{equation}}

\begin{document}

\title{Superinflation, quintessence, and the avoidance of the initial 
singularity}

\date{\today}\author { A. Saa$^{1,2,3}$, E. Gunzig$^{1,4}$, L. Brenig$^{1,5}$,
V. Faraoni$^{1,6}$, T.M. Rocha Filho$^{7}$ and A.
Figueiredo$^{7}$.}
\address {$^1$ RggR, Universit\'e Libre de Bruxelles,
CP 231, 1050 Bruxelles, Belgium}
\address{$^2$ Dep.  F\'\i sica Fonamental, Universitat de Barcelona,
Av. Diagonal 647, 08028 Barcelona, Spain }
\address{$^3$ Dep.  Matem\'atica Aplicada, IMECC--UNICAMP, CP 6065,
13081-970 Campinas, SP, Brazil }
\address{$^4$ Instituts Internationaux de Chimie et de Physique Solvay,
CP 231, 1050 Bruxelles, Belgium}
\address{$^5$ Service de Physique Statistique, Universit\'e Libre de
Bruxelles, CP 231, 1050 Bruxelles, Belgium}
\address{$^6$ INFN-Laboratori Nazionali di Frascati, Box 13, 00044
Frascati, Roma, Italy}
\address{$^7$ Instituto de F\'\i sica, Universidade de Brasilia, 70.910-900
Brasilia, DF, Brazil}    

\maketitle

\begin{abstract}
We consider the dynamics of a spatially flat
universe dominated by a self-interacting 
nonminimally coupled scalar field. The structure of the phase
space and complete phase portraits for the conformal
coupling case are given. It is shown that the non-minimal coupling
modifies drastically the dynamics of the universe.
New cosmological behaviors are identified,
including superinflation ($\dot{H}>0$), avoidance of big bang singularities
through classical birth of the universe from empty Minkowski space, and
spontaneous entry
into and exit from inflation. The relevance of this model to
the description of quintessence is discussed.
\end{abstract}
\vspace{0.5cm}


\section{Introduction}

The description of the matter content of the cosmos with a single scalar 
field is appropriate during
important epochs of the history of the universe \cite{KolbTurner}.
In this article, 
a dynamical system approach to a self-consistent nonsingular cosmological
history is presented in the framework of the classical Einstein equations 
with a nonminimally coupled scalar field.
The complete structure of the phase portrait and of the dynamical
behavior is presented for the case of a scalar field 
conformally coupled to the spacetime curvature and with
a quartic self-interaction potential. This exhaustive analysis
is made possible thanks to a reduction of the dynamics to
a two-dimensional manifold embedded in the original three-dimensional
phase space, a general property show earlier by some of the authors
\cite{IJTP1} for a classical scalar field in a spatially flat
universe with arbitrary self-interaction potentials and arbitrary
coupling to the curvature.
Solutions with special dynamical
interest are identified, namely heteroclinic and homoclinic
solutions in the reduced two-dimensional phase-space, and
their relevance to a possible classical birth of the universe from empty 
space is discussed. We recall that heteroclinic  
trajectories in a phase space correspond to bounded
solutions connecting two different fixed points. Typically, they
play the role of separatrices, determining regions of the
phase space with qualitative
distinct dynamical behaviors. Homoclinic trajectories, on the other
hand, correspond to solutions starting and ending at the same
fixed point. Their relevance to chaotic motions has been intensively
discussed in the literature\cite{Ozorio}. Despite the fact that the
model presented here has no chaotic behavior, its homoclinic solutions
will probably mark candidate regions of the phase space for chaotic
motions if a small perturbation to the equations is introduced.

Our model consists in a universe filled with a self interacting 
non-minimally coupled scalar field.
A crucial ingredient of the physics of scalar
fields in curved spaces is their nonminimal coupling to the scalar
curvature $R$ of spacetime, which is required by first loop corrections
\cite{loop},  
by specific particle theories \cite{preprint}, 
and by scale-invariance arguments at the classical level
\cite{CCJ}. It is well known that
nonminimal coupling dictates the success or failure of inflationary
models \cite{Abbott}; more generally, it turns out to
strongly affect the cosmic dynamics, which is qualitatively richer than 
in the minimally coupled case.  We show, indeed, that nonminimal coupling
leads to new dynamical behaviors, such as a regime that we propose to call
{\em superinflation} ($\dot{H}>0$), which cannot be achieved with minimal
coupling \cite{LiddleParsonsBarrow}, and spontaneous entry into
and exit from inflation, with or without a cosmological constant. 
Spontaneous superinflation provides a classical
alternative to semiclassical birth of the universe from empty Minkowski
space \cite{GunzigNardone,IJTP1}, which is
impossible with minimal coupling.

In the next section, we review our model, 
recently proposed in \cite{PRD}, and give some definitions.
The aspect of the phase portraits are presented in section \ref{phase}.
Section \ref{asympt} is devoted to the asymptotic analyses
of some especial solutions. The last section contains the 
concluding remarks.

\section{The model}

We consider the nonminimally coupled theory described by the action
\begin{equation}   \label{action}
S=\frac{1}{2} \int d^4x\sqrt{-g}\left(-\, \frac{R}{\kappa } 
+g^{\mu \nu}\partial_\mu \psi \partial_\nu \psi 
-2V +\xi R\psi ^2  \right) \; ,
\end{equation}
where $R$ denotes the scalar  curvature, 
$\psi$ is the scalar field,
$\kappa \equiv 8\pi G$ ($G$ being Newton's constant), and $\xi $ is the
nonminimal coupling constant. A cosmological constant $\Lambda$, if
present, is incorporated in the scalar field potential $V(\psi )$.
We use the full conserved scalar field stress-energy tensor 
\begin{equation}   \label{stressenergy}
T_{\mu\nu} =  \partial_{\mu}\psi \partial_{\nu} \psi-\xi \left(
\nabla_{\mu}\nabla_{\nu}
-g_{\mu\nu} \Box \right)( \psi^2 )+\xi G_{\mu\nu} \psi^2 
 -\frac{1}{2} g_{\mu\nu} \left( \partial_{\alpha} \psi 
\partial^{\alpha} \psi -2V( \psi) \right) 
\end{equation}
(where $G_{\mu\nu}$ is the Einstein tensor), thereby 
avoiding the widespread effective coupling
$\kappa_{\rm eff}=\kappa \left( 1-\kappa \xi\psi^2 \right)^{-1}$
in the Einstein equations
$G_{\mu\nu} =\kappa T_{\mu\nu}$.  
We consider here 
the dynamics of a spatially flat Friedmann-Robertson-Walker universe
with line element $ds^2=d\tau^2-a^2( \tau) \left(  dx^2+dy^2+dz^2
\right)$. This yields the trace equation $R=-\kappa \left(
\sigma-3p\right)$, the energy constraint $3H^2=\kappa\sigma$ (which
guarantees that the energy density $\sigma \geq 0$), and the
Klein-Gordon equation. More explicitly,
\begin{equation}  \label{trace}
6\left[ 1 -\xi \left( 1- 6\xi \right) \kappa \psi^2
\right] \left( \dot{H} +2H^2 \right) 
-\kappa \left( 6\xi -1 \right) \dot{\psi}^2  
- 4 \kappa V  + 6\kappa \xi \psi \frac{dV}{d\psi} = 0 \; , 
\end{equation}
\begin{equation} \label{Hamiltonianconstraint}
\frac{\kappa}{2}\,\dot{\psi}^2 + 6\xi\kappa H\psi\dot{\psi}
- 3H^2 \left( 1-\kappa \xi \psi^2 \right) + \kappa  V =0 \, ,
\end{equation}
\be    \label{KleinGordon}
\ddot{\psi}+3H\dot{\psi}+\xi R \psi +\frac{dV}{d\psi} =0 \; .
\ee
The (time-dependent) equation of state of the $\psi$ field, rather than
being imposed {\em
a priori}, follows self-consistently from the dynamics.
 In the trace equation (\ref{trace}), the
second derivative $\ddot{\psi}$ 
that appears in the pressure has been replaced by its expression
given by the Klein-Gordon equation.  Clearly, in the set of equations 
(\ref{trace})-(\ref{KleinGordon}), the sub-system
(\ref{trace})-(\ref{Hamiltonianconstraint}) is a closed 
implicit two-dimensional system for $\psi$ and $H=\dot{a}/a$
(note that this dimensional reduction is
not possible for spatially curved universes \cite{ALO,IJTP1}). 
After solving these implicit equations for $\dot{\psi}$ and 
$\dot{H}$, one has
\begin{equation}  \label{psidot}
\dot{\psi}=-6 \xi H \psi \pm\frac{1}{2\kappa} \sqrt{{\cal G} ( H, \psi
)}
\; ,
\end{equation}
\begin{equation} \label{Hdot}
\dot{H}= \frac{1}{1+ \kappa \xi ( 6\xi -1 ) \psi^2 } 
\left[ 3 \left( 2\xi -1 \right) H^2 
+ 3 \xi ( 6\xi -1 )
( 4\xi -1)\kappa H^2 \psi^2 \mp \xi
( 6\xi -1) H \psi \sqrt{{\cal G}} 
+\left( 1-2\xi \right) \kappa V ( \psi ) -\kappa \xi \psi
\frac{dV}{d\psi} \right] \; ,    
\end{equation}
where
\be  \label{G}
{\cal G} ( H, \psi ) = 8 \kappa^2 \left[ \frac{3H^2}{ \kappa} -V ( \psi )
+3 
\xi ( 6\xi -1 ) H^2 \psi^2  \right] \; .
\ee
Due to the energy constraint (\ref{Hamiltonianconstraint}), the
trajectories are restricted 
to a two-dimensional manifold $\Sigma$ in the three-dimensional
$\left( H, \psi, \dot{\psi} \right) $ 
original phase space, possibly with ``holes''
(dynamically forbidden regions) 
corresponding to ${\cal G}(H,\psi)<0 $ (cf. Eq.~(\ref{psidot})). $\Sigma$
is composed of two sheets corresponding to the positive or negative sign
in Eq.~(\ref{psidot}). The
two sheets 
smoothly
join on the boundary ${\cal G}=0$ of the dynamically
forbidden region. In the present work, we restrict to the potential
\be  \label{potential}
V( \psi ) = \frac{3\alpha}{\kappa} \psi^2 -\frac{\Omega}{4} \psi^4 -
\frac{9\omega}{\kappa^2} \;,
\ee
consisting of a mass term, a quartic self-coupling and, possibly, a
cosmological constant term. For consistency with
previous works \cite{IJTP1} we use the symbols $\alpha \equiv \kappa
m^2/6$ ($m$ being the scalar field mass) and $\omega \equiv
-\kappa^2\Lambda/9 $. Figures 1 and 2 present some aspects
of $\Sigma$ for the potential (\ref{potential}). However, several of our
main results do not depend on the details of $V(\psi)$.
\begin{center}
\begin{figure}[h]
\label{sheet}
\epsfxsize=8.0cm
\epsfbox{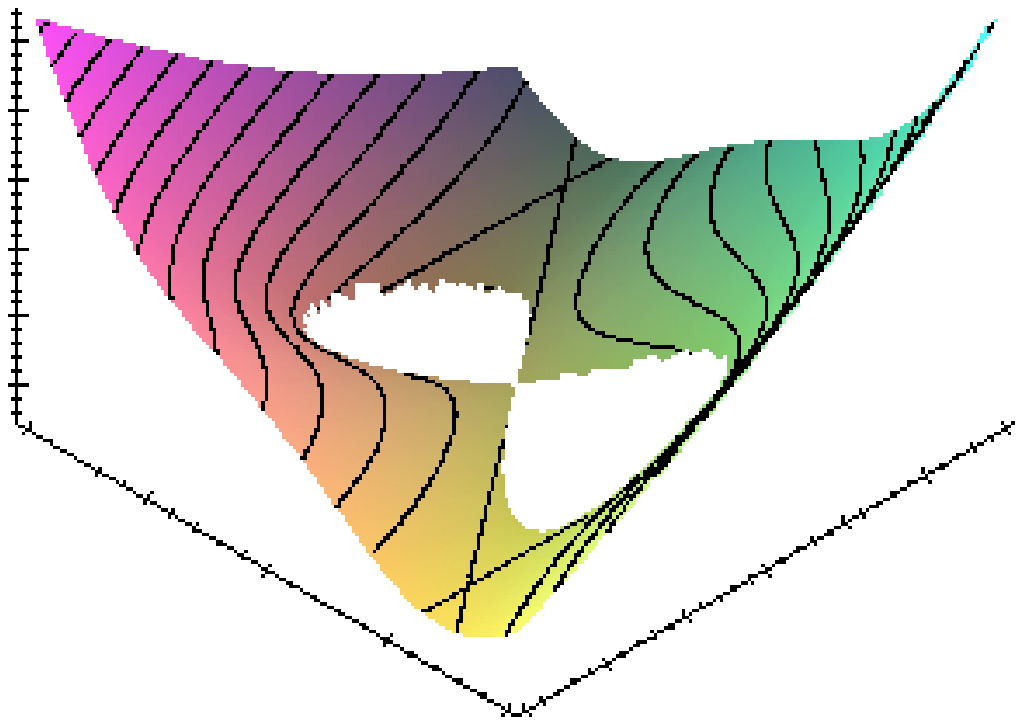}

\epsfxsize=8.0cm
\epsfbox{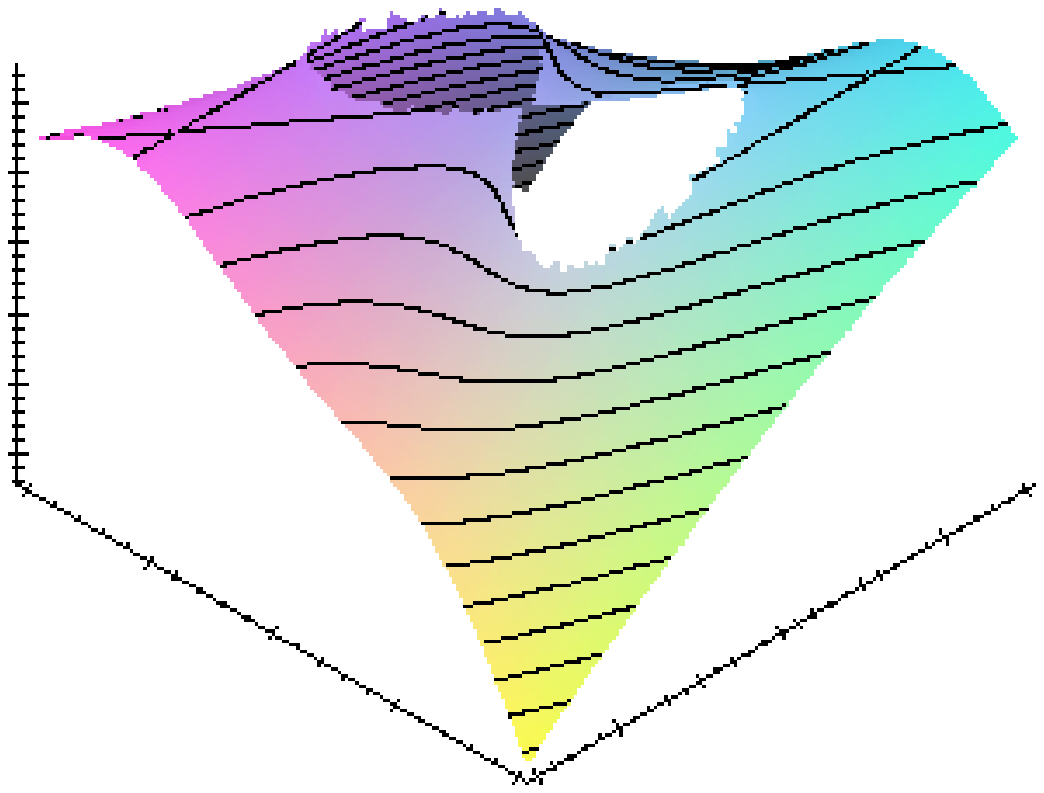}
\vspace{0.5cm}
\caption{Aspect of the two-dimensional manifold $\Sigma$, 
embedded in the three-dimensional phase space
$(\psi,H,\dot{\psi})$, for the
potential (\ref{potential}). The upper graph represents the
``+'' sheet, while the lower one depicts the ``-'' sheet.
In fact, they are not disconnected, 
the two sheets join smoothly on 
the boundary ${\cal G}$=0 of the dynamically forbidden region,
corresponding to the showed ``holes''. The lines of constant
$\cal G$ are drawn on the sheets.}
\end{figure}
\end{center}

\begin{center}
\begin{figure}[h]
\label{cone}
\epsfbox{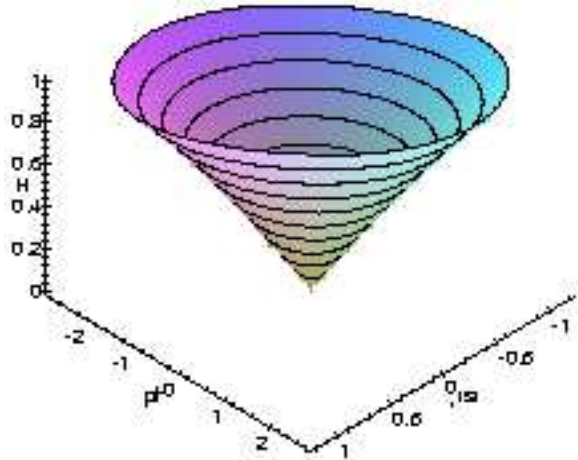}
\vspace{0.5cm}
\caption{Zoom,  near the origin, of the two-dimensional manifold $\Sigma$,
embedded in the three-dimensional phase space
$(\psi,\dot{\psi},H)$,  for the
potential (\ref{potential}). The two sheets join
forming two symmetric cones with their apex at the origin. Only the
cone corresponding to $H>0$ is presented here, and the lines of
constant $H$ are drawn.}
\end{figure}
\end{center}

\section{Phase space portraits}
\label{phase}

In the following, for simplicity, we project the dynamics of the phase
space onto the $(H, \psi )$ plane, but the true  
nature of $\Sigma$ should always be kept in mind. The fixed 
points of the system (\ref{trace})-(\ref{KleinGordon}) include
de Sitter solutions with constant scalar field
\be    \label{fixedpoints}  
H^2_0=\frac{ 3(\alpha^2-\Omega\omega)}{\kappa ( \Omega - 6\xi\alpha )} \;,
\;\;
\psi^2_0=\frac{ 6(\alpha-6\xi\omega)}{\kappa ( \Omega - 6\xi\alpha )} 
\ee
$\left( \Omega\neq 6\alpha\xi \right)$, and the solutions 
$\left( H,\psi \right)=\left( \pm \sqrt{-3\omega/\kappa}, 0 \right)$.
The fixed points  (\ref{fixedpoints})  exist also
for $\omega=0$, due to the presence of the matter field $\psi$ (in this
case, the two points $\left( \pm \sqrt{-3\omega/\kappa}, 0 \right)$
collapse into the Minkowski space fixed point, which 
is at the apex of the cones described in Figure 2). Here,
we restrict ourselves to the case of conformal coupling, $\xi=1/6$.

The function 
\be  \label{lyapunov}
L( \psi, \dot{\psi})= \frac{1}{2} \dot{\psi}^2 + \frac{\alpha}{4} \,
\psi^4 - \frac{3\omega}{\kappa} \, \psi^2 + V( \psi) 
\ee
is such that $dL/dt =-3H\dot{\psi}^2$ along the trajectories. For $H>0$,
$L$ is a Lyapunov function  in a region containing 
the origin; the solutions are then confined  by closed lines of constant
$L$, 
implying asymptotic convergence to the fixed points on the $H$ axis (from
Eq.~(\ref{Hamiltonianconstraint}), if $\psi $ and $\dot{\psi}$ vanish, $H$
goes to
$V(0)$). This behavior is confirmed by exhaustive numerical 
simulations\cite{PRD} and 
reported in the following.
We first exclude a cosmological constant
by setting $\omega=0$. The phase portrait qualitatively differs according
to
the ratio $\Omega/\alpha$.

\subsection{ The case $\Omega=2\alpha$} 

 The Minkowski space   $\left( H, \psi,
\dot{\psi} \right)=\left( 0, 0, 0
\right) $ is a fixed point, attractive for $H>0$ and repulsive for $H<0$;
the projections of the de Sitter spaces 
$\left( \pm H_0, \pm \psi_0, 0 \right) $ are saddle points, i.e. they
possess attractive and repulsive eigendirections in the phase space
(Fig.~3a).
They are of two kinds: expanding ($H\psi>0$) or contracting ($H\psi<0$).
The following solutions, present only in this 
particular case, 
\be  \label{heteroclinic}
H( \tau ) =\sqrt{\frac{C}{2}} \tanh \left( \sqrt{2C}\, \tau \right)
\;,\;\;\;\;\;\;
\psi=\pm \psi_0 \equiv \pm \sqrt{\frac{6}{\kappa}} 
\ee
(where $C=\dot{H}+2H^2 = -R/6 $ is constant), correspond to heteroclinic
straight
lines connecting de Sitter fixed points, starting along
the repulsive eigendirection of one of them and ending along the
attractive eigendirection of the other (Fig.~3a). They are
tangent to the boundary of the forbidden regions at  
$\left( H, \psi \right) =\left( 0, \pm\psi_0 \right) $.
For $\left| H \right| >\sqrt{C/2}$, another straight line solution is
obtained from the general form
\be  \label{exact}
H( \tau)=\sqrt{\frac{C}{2}}\, \frac{w_1 {\mbox e}^{\sqrt{C/2}\tau}-
w_2 {\mbox e}^{-\sqrt{C/2}\tau}}
{w_1 {\mbox e}^{\sqrt{C/2}\tau}+
w_2 {\mbox e}^{-\sqrt{C/2}\tau}}\; , \psi=\psi_0 \; ,
\ee
where $w_1$ and $w_2$ are integration constants.
The nonsingular solutions (\ref{heteroclinic}) connect a
contracting ($\tau \rightarrow -\infty$) de Sitter regime to a minimum
nonvanishing value of the scale factor ($\tau=0$), and then to an
expanding de Sitter regime ($\tau \rightarrow +\infty$).

In addition to these straight lines, we found numerically 
other heteroclinic solutions: one starting at $\left(
H,\psi \right)=( 0,0) $ and ending at  
$( -H_0, \psi_0) $, and another one from 
$\left( H_0,\psi_0 \right)$ to $( 0,0) $. A third solution starts at
the expanding de Sitter point and goes to infinity, while another one
comes from infinity and arrives to the contracting de Sitter point. The
phase portrait is symmetric about the origin.

Near the fixed point $\left(
0,0,0 \right)$, numerical analysis confirms the peculiar behavior
suggested
by the Lyapunov function: orbits approaching this point with positive
$H$ are attracted to it, bouncing back and forth infinitely many times off
the
${\cal G}=0$ boundary in the $\left( H, \psi \right) $ projection
(Fig.~3a). In the space $\left( H,
\psi,
\dot{\psi} \right) $ these orbits are seen to spiral down on a cone
towards its apex at the origin.  
The cone results from the union of the two sheets in the vicinity of the
origin, as it is shown in Figure 3. 
Along the spiral, the orbit passes almost periodically from one
sheet to the other with period $ \tau_{\rm bounce}=2\pi/m$ ($\tau_{\rm bounce}$ is
obtained from the asymptotic analysis of the next section. 
Typically, after a few bounces, the period coincides with
$\tau_{\rm bounce} $ with good accuracy). A similar 
behavior for $\Omega <0$ was reported in the earlier numerical analysis
of \cite{ALO}, but  using the effective coupling $\kappa_{\rm eff}( \tau)$ and
the variables $\psi$ and $\dot{\psi}$.

\begin{center}
\begin{figure}
\epsfxsize=12.0cm
\epsfbox{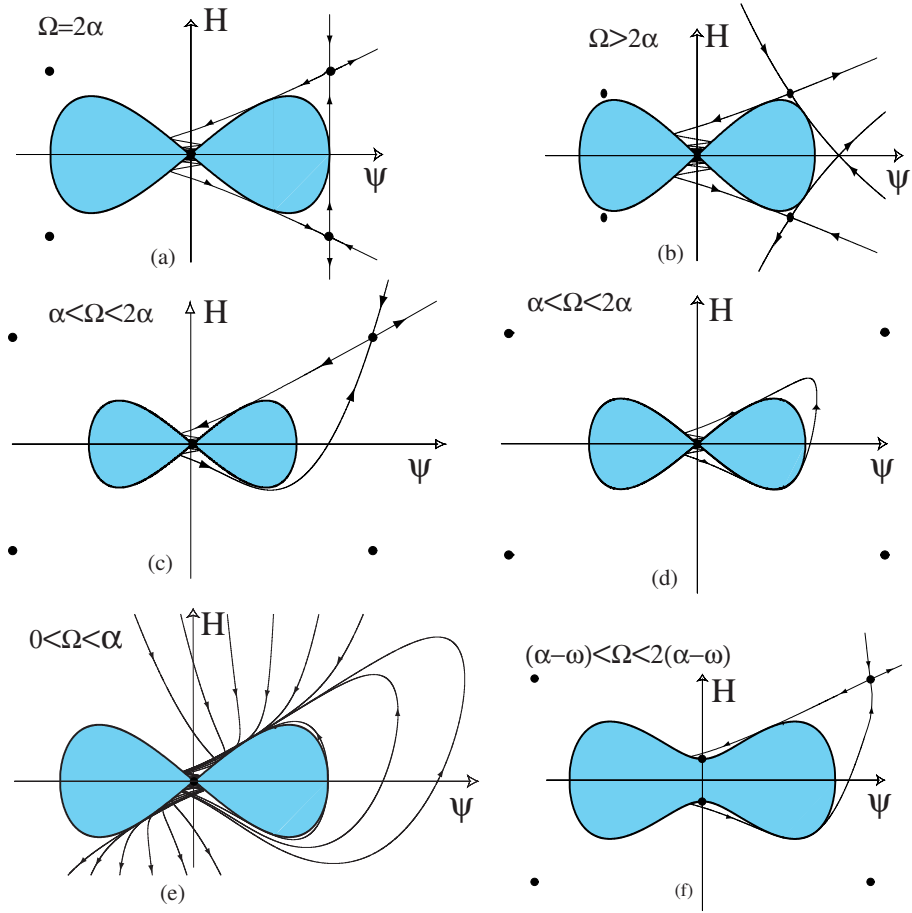}
\vspace{0.5cm}
\caption{Qualitative phase portraits for the system  
(\ref{trace})-(\ref{KleinGordon}). Shadowed regions correspond to the
dynamically forbidden regions (${\cal G}(H,\psi)<0 $, cf. Eq.~(\ref{psidot})).
Figures a-e, were obtained by using $\Omega/\alpha$, respectively,
2, 5, 3/2, 3/2, and 1/2, and $\omega=0$. Figure f corresponds to
the case $\omega=1/10$ and $\Omega/(\alpha-\omega)=3/2$.}
\end{figure}
\end{center}

In the $H<0$ half-plane, the situation is reversed: orbits starting with
$H<0$ are repelled by the origin and depart from it bouncing
off the ${\cal G}=0$ boundary.

\subsection{The case $\Omega> 2\alpha$} 

The situation (Fig.~3b) is analogous to
the previous one, but now the straight heteroclinic solutions are
missing, and are replaced by the solution starting at the contracting
de Sitter fixed point and escaping to infinity, and by the solution coming
from infinity and arriving to the expanding de Sitter point. The
two-sheeted structure of $\Sigma$ implies that no actual intersections
occur between different orbits in Fig.~3b, which live in different
sheets but are projected on the same plane.

\subsection{The case $\alpha < \Omega< 2 \alpha $} 

As shown in Fig.~3c, there
are no straight heteroclinic lines but
new interesting features emerge. A new heteroclinic solution appears
starting from the origin and ending in the expanding de Sitter fixed
point. As in the previous case, the quadrant $\psi>0$, $H<0$ is obtained
from the $\psi>0$, $H>0$ one by reflection about the $\psi$-axis and
time-reversal in Fig.~3c.

The crucial feature of this case is the appearance of a dense set of
homoclinic solutions (Fig.~3d) departing from the origin with negative $H$
and
returning to it with positive $H$, going around the forbidden region.
Superinflation plays a central role along these orbits; only a regime with
$\dot{H}>0$ permits the smooth transition from an initial contracting
($H<0$) phase to an expanding one ($H>0$). This transition occurs at the
nonvanishing minimum of the scale factor. The behavior of this family of
homoclinics, as well as of the other solutions, is universal: they rapidly
converge in the spiraling region near the origin, irrespective of 
initial conditions. Since all these homoclinics originate from
a neighborhood of the Minkowski fixed point 
due to its own instability with respect to perturbations with $H<0$,
and come back to that point, due to the stability for perturbations with
$H>0$, this 
behavior constitute a classical alternative to the previously proposed
semiclassical birth of the universe from empty space
\cite{GunzigNardone,IJTP1}.

\subsection{The case $0<\Omega< \alpha$} 

The fixed points (\ref{fixedpoints})
disappear and the only bounded solutions are the
homoclinics
associated with the origin (see Fig.~3e). This situation is therefore the
most favorable for the classical spontaneous exit from empty Minkowski
space.

\subsection{The case $\omega < 0$}

Analogous results hold if a small cosmological constant is present (see
Figs.~3f), with the phase portrait being classified according to
$\Omega / ( \alpha-\omega )$, but the fixed point $(0,0,0)$ of the
$\omega=0$ case splits into two de Sitter fixed points with memory of
the previous stability properties. Now, the approximate period between two
consecutive bounces is 
\be
\tau_{\rm bounce}=\frac{2\pi}{\sqrt{m^2+ \frac{3\omega}{4\kappa}}} \; .
\ee

\section{Asymptotic analysis}
\label{asympt}

For the conformally coupled case, the trace and the Klein Gordon equations
(\ref{trace}) and (\ref{KleinGordon}), after a rescaling and
the explicit substitution of the Hamiltonian
constraint (\ref{Hamiltonianconstraint}), read
\begin{eqnarray}
\label{eee}
  & \dot{H} + 2H^2 - \alpha\psi^2 + 6\omega = 0 & \nonumber \\
  & \ddot{\psi} + 3H\dot{\psi} + 6(\alpha -\omega)\psi - 
  (\Omega-\alpha)\psi^3 = 0 &  
\end{eqnarray}
We have special interest in the asymptotic behavior of the
solutions of (\ref{eee}) in the vicinity of the attractive
fixed points. Let us start with the case $\omega=0$. In this
case, our region of interest is the neighborhood with
$H>0$ of the origin. We search for asymptotic solutions of
the form
\begin{eqnarray}
\label{asy}
\psi(t) &=& \frac{f_1(t)}{t} + \frac{f_2(t)}{t^2} + {\cal O}(t^{-3})
\nonumber \\
H(t) &=& \frac{g_1(t)}{t} + \frac{g_2(t)}{t^2} + {\cal O}(t^{-3})
\end{eqnarray}
with $f_1(t),f_2(t),g_1(t),$ and $g_2(t)$ bounded for large $t$.
Inserting (\ref{asy}) in the equations (\ref{eee}) one has
\begin{eqnarray}
\label{asy2}
\frac{1}{t}\left(f''_1+6\alpha f_1 \right) +
\frac{1}{t^2}\left( 
f''_2 - 2f'_1 + 3g_1f'_1 + 6\alpha f_2
\right) &=& {\cal O}(t^{-3})
\nonumber \\
\frac{1}{t}g'_1 +
\frac{1}{t^2}\left( g'_2 - g_1 + 2g^2_1 - \alpha f_1^2  \right) &=& 
{\cal O}(t^{-3})
\end{eqnarray}
By demanding the exactness of (\ref{asy2}) up to $t^{-3}$ order,
one gets immediately that $g_1$ is a constant and
$f_1 = A\cos(\sqrt{6\alpha}t + \delta)$, leading to
\begin{equation}
\label{asy3}
f''_2 +6\alpha f_2 = A\sqrt{6\alpha}(3g_1 - 2)\sin
(\sqrt{6\alpha}t + \delta), 
\end{equation}
which has the general solution
\be
f_2 = B\cos(\sqrt{6\alpha}t + \theta) - \frac{A(3g_1-2)}{2\sqrt{6\alpha}}
\left( \cos(\sqrt{6\alpha} t + \delta) \sqrt{6\alpha}t  -
\cos(\sqrt{6\alpha}t)\sin\delta\right).
\ee
In order to guarantee the boundedness of $f_2$, it is necessary to have
$g_1 = 2/3$. From the equation corresponding to the $t^{-2}$ term
in the trace equation, we have
\begin{equation}
g'_2 = A^2\alpha\cos^2(\sqrt{6\alpha}t +\delta) - \frac{2}{9},
\end{equation}
which is solved by
\be
g_2 = \frac{A^2\alpha}{2\sqrt{6\alpha}} \left(
\cos(\sqrt{6\alpha}t +\delta)\sin(\sqrt{6\alpha}t +\delta) +\delta
\right) +
\left(\frac{A^2\alpha}{2} - \frac{2}{9}\right)t.
\ee
Again, by requiring the boundedness of $g_2$ for large $t$, we
get the condition $A=2/(3\sqrt{\alpha})$. We have, finally, the following asymptotic
solution for the $\omega=0$ case
\begin{eqnarray}
\label{asy5}
\psi(t) &=& \frac{2\cos\sqrt{6\alpha}t}{3\sqrt{\alpha}t} + {\cal O}(t^{-2})
\nonumber \\
H(t) &=& \frac{2}{3t} + {\cal O}(t^{-2}).
\end{eqnarray}
{}From (\ref{asy5}) we can obtain the characteristic 
period $\tau_{\rm \rm bounce} = 2\pi/\sqrt{6\alpha}$. 
Also, the asymptotic solution  
(\ref{asy5})
implies that all solutions ending in the origin will behave
as matter dominated universes for large $t$, corresponding
to $a( t) \propto t^{2/3}$.

The case with a small cosmological constant can be treated
analogously, taking into account that now the relevant fixed
point is $\psi=0$ and $H=\sqrt{-3\omega}$. The search for
asymptotic solutions of
the form
\begin{eqnarray}
\label{asy7}
\psi(t) &=& \frac{f_1(t)}{t} + \frac{f_2(t)}{t^2} + {\cal O}(t^{-3})
\nonumber \\
H(t) &=& g_0(t) + \frac{g_1(t)}{t} + \frac{g_2(t)}{t^2} + {\cal O}(t^{-3})
\end{eqnarray}
with $f_1(t),f_2(t), g_0(t),  g_1(t),$ and $g_2(t)$ bounded for large $t$,
as before, give rise to the equations
\begin{eqnarray}
\label{asy22}
\frac{1}{t}\left(f''_1+3g_0f'_1+6(\alpha-\omega) f_1 \right) +
\frac{1}{t^2}\left( 
f''_2 - 2f'_1 + 3(g_1f'_1+g_0f'_2-g_0f_1) + 6(\alpha-\omega) f_2
\right) &=& {\cal O}(t^{-3}),
\nonumber \\
\left(g'_0 + 2g_0 + 6\omega\right)  + 
\frac{1}{t}\left(g'_1 + 4g_0g_1\right) +
\frac{1}{t^2}\left( g'_2 - g_1 +4g_0g_1+ 2g^2_1 - \alpha f_1^2  \right) &=& 
{\cal O}(t^{-3}).
\end{eqnarray}
Our asymptotic equations now are considerably more complicated,
but, nevertheless, we can obtain an unambiguous period 
$\tau_{\rm bounce}$. From (\ref{asy22}), one has that $g_0$ decreases
exponentially to the value $\sqrt{-3\omega}$ for large $t$, implying
that $f_1$ converge to the form 
$f_1 = A\exp({-3\sqrt{-3\omega}t/2)\cos(\sqrt{6\alpha + 3\omega/4}t+\delta)}$,
{}from which one can obtain 
$\tau_{\rm bounce} = 2\pi/\sqrt{6\alpha + 3\omega/4}$. The scale factor,
in this case, obeys $a(t) \propto \exp{\sqrt{-3\omega}t}$ for large $t$, 
as in the de Sitter universe.

\section{Discussions}

The $ \left( H, \psi \right)$ plane is divided into sectors by the
straight lines $H=\pm \sqrt{\alpha/2} \, \psi$, 
$H=\pm \sqrt{\alpha} \, \psi$
$H=\pm \sqrt{2\alpha} \, \psi$ corresponding, respectively, to
$\dot{H}=0$, $\ddot{a}=0$, and pressure $p=0$ (the $H$-axis  corresponds
to $p=\sigma/3$). The lines  $H=\pm \sqrt{\alpha/2} \, \psi$ mark the
transition between inflationary ($\ddot{a}>0$ and $\dot{H} \leq 0$) and
superinflationary ($\dot{H}>0$)  regimes. The lines 
$H=\pm \sqrt{\alpha} \, \psi$ divide regions corresponding to inflation
and to decelerated expansion, while 
$H=\pm \sqrt{2\alpha} \, \psi$ divide regions of positive and negative
pressures (Fig.~4). The crucial condition for superinflation to occur is
that  the line $\dot{H}=0$ (or parts of it) should belong
to the dynamically accessible region ${\cal G}\geq 0$ of the $\left( H,
\psi \right) $ plane. This implies that, for an arbitrary potential $V(
\psi)$,
superinflation corresponds to $\psi dV/d\psi \leq 0 $ (this result will
be discussed in detail elsewhere). For our particular potential
(\ref{potential}), this requires $\Omega>0$. 
The superinflationary behavior occurs only once along each homoclinic and
brings the solution from the primordial Minkowski neighborhood to the
succession of eras corresponding to different equations of state (during
each bounce), towards infinite dilution and equation of state $p=0$. In
fact, asymptotic analysis for $\tau \rightarrow +\infty$ and for any
value of $\alpha$ and $\Omega$ (with $\omega=0$) shows that the scale
factor $a( \tau ) $ exhibits oscillations of concavity corresponding to
accelerated and decelerated epochs. These oscillations are damped as $\tau
\rightarrow +\infty$; in this regime, $a( \tau) \propto \tau^{2/3}$ and
the universe becomes matter dominated.

\begin{center}
\begin{figure}
\epsfxsize=12.0cm
\epsfbox{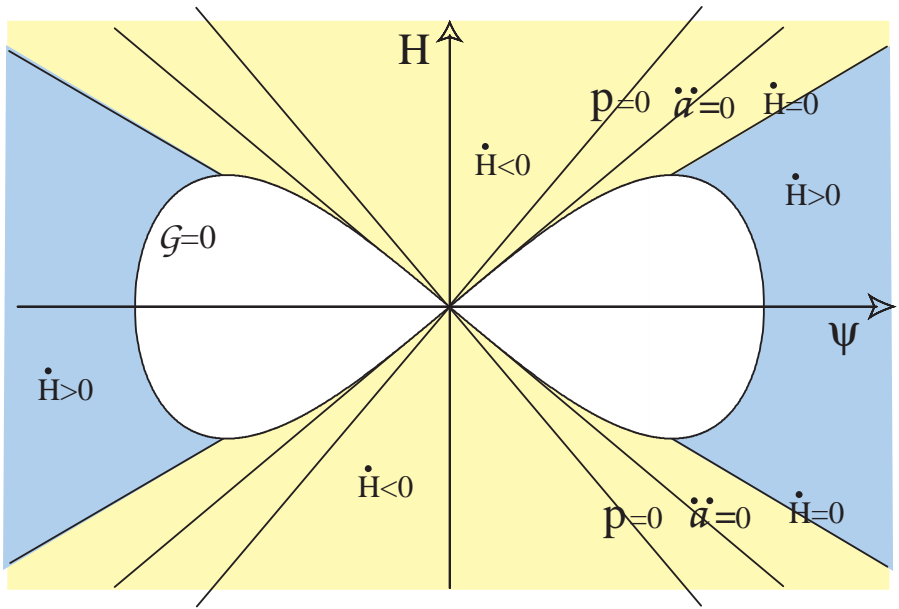}
\vspace{0.5cm}
\caption{The plane $(H,\psi)$ for the
$\omega=0$ case and the equation of state for $\psi$. The darkest
region corresponds to the superinflation regime ($\dot{H}>0$).
During the bounces of the $\psi$ solution in the region $\dot{H}<0$, its
equation of state corresponds to, respectively, 
radiation domination (crossing the $H$-axis),
matter domination ($p=0$), and re-acceleration (between ${\cal G}=0$
and $\ddot{a}=0$ lines.
These oscillations are damped as $\tau
\rightarrow +\infty$, 
the universe becomes matter dominated $a( \tau) \propto \tau^{2/3}$ and
tends to infinite dilution.}
\end{figure}
\end{center}

While it is not claimed here that the evolution of our universe is
modeled by an entire orbit of the system 
(\ref{trace})-(\ref{KleinGordon}) on the accessible manifold  
$\Sigma$, the application to specific eras of the cosmological history is
intriguing.
Indeed, in the bounces reported above (during which $\dot{H}<0$), one
encounters, respectively, radiation domination crossing the $H$-axis,
matter domination ($p=0$), acceleration (a possible quintessence model?)
until the next bounce in the $\left( H, \psi \right)$ projection, where
this sequence is reversed.

If we identify one period as our cosmological history, then the reported
accelerated expansion of the universe today \cite{SN} suggests to locate
our epoch in the sector between the line $H=\sqrt{\alpha} \, \psi$ and the
${\cal G}=0$ boundary. The identification of the age of the universe
($\sim 10^{17}$~s) with
$\tau_{\rm bounce}$ would then yield the scalar field mass $m \simeq
10^{-13}$~eV, which is suggestive of an axion \cite{KolbTurner} or of an 
ultralight pseudo-Goldstone boson (that the quintessence field should be
very light was already suggested \cite{Binetruy}).

As a step towards a realistic model, one can include a second scalar field
coupled to $\psi$, which has the meaning of a fundamental field
(as is done, e.g., in hybrid inflation),  or mimics a
baryonic or other fluid. In spite of the higher dimensionality of
the phase space, many of the features exposed here for a single scalar
field survive \cite{twoscalars}.
Although very simple, we think our classical model opens interesting
avenues to the understanding of quintessence in terms of a non-minimally
coupled scalar field.

Finally, one should keep in mind that our
detailed semi-analytical analyses are possible thanks to the
reduction of the original three-dimensional system to a two-dimensional
one on a smooth manifold $\Sigma$, which, in turn, strongly indicates
the absence of chaos (confirmed by an exhaustive numerical analysis).
Generically, such a reduction is not possible for the case of 
a spatially curved spacetime and/or several matter fields non-minimally
coupled to the curvature, despite of many similar results.
This is object of present investigations\cite{twoscalars}.

\acknowledgements

We acknowledge the Centre de Calcul Symbolique sur Ordinateur for
the use of computer facilities, and financial support from the EEC
(grant HPHA-CT-2000-00015), from OLAM, Fondation pour la Recherche
Fondamentale (Brussels), from FAPESP (S\~ao Paulo, Brazil), and from the
ESPRIT Working Group CATHODE.

\end{document}